\documentclass[a4paper,11pt]{article}
\usepackage{latexsym}
\usepackage{feynmf}		
\def\ba{\begin{eqnarray}}
\def\ea{\end{eqnarray}}

\def\lb{\label}
\def\be{\begin{equation}}
\def\ee{\end{equation}}

\begin{document}
\title{A common scenario leading to a small vacuum energy and stable supermassive particles}
\author{Osvaldo P. Santill\'an \thanks{Departamento de Matem\'atica, FCEyN, Universidad de Buenos Aires, Buenos Aires, Argentina
firenzecita@hotmail.com and osantil@dm.uba.ar.}}
\date {}
\maketitle
\begin{abstract}
A toy model giving rise to long lived super heavy particles and an small vacuum density energy, of the order of the one measured in the present universe, is constructed. This model consists in hidden sector invariant under an $SU(2)_L$ gauge symmetry, whose masses are provided by the standard Higgs mechanism. It is assumed that the standard model particles are also charged under this interaction. The hidden fermions and the hidden Higgs are super-heavy, which mass values close to the GUT scale. In addition, there is an spontaneously broken $U(1)$ chiral symmetry, giving rise to a pseudoscalar Goldstone boson which we refer as a "hidden axion".
We model the vacuum energy of the universe as the potential energy of this pseudoscalar, and this fixes several scales of the model. In particular, it is shown that the interaction between the hidden and the ordinary sector is very weak, of the order of the gravitational one. The approach to the vacuum energy problem presented here is a quintessence like mechanism, in which it is assumed that the true vacuum density energy is zero for some unknown reason, except for the contribution of the light axion. As consequence of the weakness of this interaction, the hidden Higgs is long lived and may act as a super heavy component of the dark matter at present times.

\end{abstract}
\section{Introduction}
The density of energy of the present universe seems very close to the critical one $\rho_c\simeq 10^{-47}$GeV \cite{Carroll}. This value is not explained by the current understanding of QFT since the scale of supersymmetry breaking and the scale of the quark condensate are around 55 and 43 orders of magnitude higher than the critical one, respectively. Thus, it is needed to explain why the vacuum density energy is so small and why it is so close to the critical one. 

An approach for explaining the smallness of the vacuum energy are the cancellation mechanisms \cite{Dolgov}-\cite{Dolgov3}. In these scenarios it is assumed the presence of an initial energy density, together with an unknown component which contributes to this density with opposite sign, in such a way that for large times the total energy density becomes very small. A different approach is suggested in \cite{Eternity}-\cite{Eternity2} where it is argued that the de Sitter space itself suffers an adiabatic catastrophe rendering it unstable. This situation was also considered in \cite{Akhmedov}. Additionally these references suggest that there are infrared effects which result in an effective screening of the vacuum energy, thus rendering it very small. Nevertheless the same technical results were obtained in \cite{Marolf} but the interpretation given there is that there is no instability. This is a topic of discussion at present.

A completely different type of models for the vacuum energy are the quintessence scenarios \cite{quintesence}. In these models the vacuum energy is associated with an slowly rolling scalar field $\varphi$ under the influence of a nearly flat potential $V(\varphi)$. The nearly flat condition insures that $\varphi$ is not at the minimum of its potential $V(\varphi)$ at the present times. Thus the vacuum energy is a temporary effect which disappears for larger times. The difference between quintessence and cancellation mechanisms is that in the former an initial energy is assumed, while in the latter no vacuum density energy is present except for the scalar field contribution.

The fact that the quintessence models do not take into account any QFT vacuum energy density from the very beginning may sound unsatisfactory for some. But it was indicated in \cite{Witten} that there are different string theories with axion like pseudoscalars \cite{axion1} miming the characteristics of the quintessence. Additionally this reference discusses several scenarios in which the true vacuum energy is zero. In this context, the idea to have only a small contribution due to a light scalar field may sound plausible. An axion pseudoscalar quintessence is in fact one of the ingredients of the present work and was also considered in \cite{quv}. Incidentally, axions play an important role in modern theoretical cosmology \cite{Wilcox} and axion emission by cosmic strings in a more theoretical setup was considered long time ago in \cite{Harari}.

 In addition to the vacuum energy problem, another topic of special interest in cosmology and particle physics is the dark matter problem. It is plausible in this context to postulate the existence of a hidden sector of particles, that is, a sector whose interaction with the known particles is weak enough to do not be detected with the current technology. This is an old concept which was introduced by Salam, Lee and Yang in order to save the parity violation \cite{Salam}-\cite{Lee} in some particle processes. At the present, there is a line of though which, in order to avoid the left right asymmetry of particle physics, postulates the existence of a hidden particle sector \cite{mirror}. Arguably, if such sectors do exist, they may contain components of the dark matter. For instance, in reference \cite{hidden33} there was analysed the possibility that dark matter is hidden, with no standard model interactions. Cosmological implications of hidden sectors has been studied for instance in \cite{hidden2} and hidden sectors with Higgs like field particles are considered in \cite{wicax}. Additionally hidden Higgs candidates for dark energy and dark matter are considered in \cite{Higgs1}-\cite{Higgs6}.

 The present work contains a hidden Higgs with a mass value of the order of $10^{13}$GeV, which is considerably larger than the ones presented in \cite{wicax}-\cite{Higgs6}. This is an extremely heavy particle. On the other hand, there is line of investigation which assumes that the dark matter contains super-heavy components \cite{review}. In a recent past, super-heavy particles were though as possible candidates to explain the origin of ultra high energy cosmic rays (UHECR) \cite{hidden3}-\cite{massiva}, i.e, cosmic rays with energy saturating the GKZ bound $10^{11}$GeV \cite{GKZ}. Some of these models assume the existence of gravitationally created super heavy particles (with mass of the order of $10^{13}$GeV) during the inflation period or topological defects originated by non thermal phase transitions. For instance the reference \cite{topsec} postulated an inflationary scenario where this defects may been originated. If objects of type are to be detected by their decays at the present or future times, they should have a time life of the order of $10^{10}$ years \cite{parecido}-\cite{Dolgui}. 
Although the AUGER project seems to discard the existence of UHECR above the GKZ bound \cite{Auger}, the study of mechanisms for long lived super massive particles may be of importance in the context of super heavy dark matter physics, even with the inclusion of supersymmetry \cite{suheavy}.

 The present work is of course, not intended to show the existence of super heavy dark matter, neither to solve the vacuum energy problem. It is focused instead on the most modest but still difficult problem of finding toy scenarios with long lived super heavy particles and with an small vacuum energy, of the order of the one measured at the present era. This is a non trivial task since, a priori, the heavier a particle is, the more probable the decay seems to be.  The scenario we will construct contains an extremely light axion field $\widetilde{a}$ under the influence of a nearly flat potential $V(\widetilde{a})$ and which reproduces the vacuum density energy of the present universe. Simultaneously, it predicts a super heavy hidden Higgs whose mean time of life is larger than the age of the universe. 
 
 We should remark that these two features mentioned above of our model are not simple "added", they are instead "linked" as follows. The axion pseudoscalar comes from a $U(1)$ symmetry breaking mechanism in a hidden sector. This sector is assumed to be invariant under a non abelian gauge symmetry and is composed by super massive particles, whose masses are provided by the standard Higgs mechanism, but with a super massive hidden Higgs. These particles are assumed to interact with the ordinary sector by the interchange of super-heavy mediating bosons of spin one, which are the gauge fields corresponding to the non abelian gauge symmetry. This implies that the standard model particles possess charges related to this interaction. The hidden axion $\widetilde{a}$  is characterized by an scale $f_{\widetilde{a}}$, which is the analogous of the pion scale $f_{\pi}$ for the hidden interaction. In addition, the requirement for the hidden axion to reproduce the correct vacuum energy $\rho_c\simeq 10^{-47}$GeV implies that it mass is extremely small, of the order of $10^{-32}$eV. These facts imply that $f_{\widetilde{a}}$ is of the order of the Planck mass $10^{19}$GeV, thus suggesting that the interaction between the hidden and the ordinary sector is of the order of the gravitational one. Due to the weakness of this hypothetical interaction, the traces of the hidden sector are not detectable by the current accelerator technology and \emph{simultaneously}, the hidden Higgs becomes long lived, with a life time larger than the estimated age of the universe.

The organization of this paper is as follows. In section 2 general properties of the classical axion model are discussed. Section 3 is based on these axion models and contains a detailed description of the hidden axion energy density and the Higgs mechanism in the hidden sector. It is also proposed an specific mechanism for the decay of the hidden Higgs into the particles of the standard model. In section 4 it is shown that the super massive Higgs is long lived with a mean time life larger than the age of the universe. Section 5 contains a discussion of the obtained results and some open perspectives are mentioned.

\section{A brief review of axion mechanisms in QCD}

The scenario to be presented is inspired in some axion models, and it will be convenient to review their main properties first.
As is well known, in ordinary QCD, the $\theta$ term associated with the instantons of the theory \cite{belavin}-\cite{thooft} 
$$
L_{\theta}=\frac{\theta}{16\pi}G^a_{\mu\nu}\widetilde{G}^{a\mu\nu},
$$
with $G^a_{\mu\nu}$ the gluon strength field, violates CP invariance when the fermions of the theory are massive. For massless QCD instead, the chiral transformation
\be\lb{rut}
\psi\to e^{i\gamma_5 \alpha}\psi,
\ee
on the fermions wave functions $\psi$ of the theory, is a classical symmetry of the lagrangian. But at the quantum level there is an anomaly in the chiral current $J_5^{\mu}$ given by
\be\lb{anomu}
\partial_{\mu}J_5^{\mu}= \frac{g^2}{16\pi}G^a_{\mu\nu}\widetilde{G}^{a\mu\nu}.
\ee
For this reason, if the fermions were massless, the chiral transformation would induce
the following one
$$
\theta\to \theta-2\alpha,
$$
to the $\theta$ parameter. This means that for massless QCD all the theories with different $\theta$ will be equivalent. It is the mass of the fermions
which spoils the chiral symmetry and simultaneously the CP invariance. 

The value of $\theta$ is not fixed by the theory itself and should be determined by the experiments. The experimental known bounds are $\theta<10^{-9}$. This value does not satisfy the majority, who considers the introduction of such small parameter is unnatural. For this reason in \cite{axion1} there was introduced an alternative to explain this naturalness problem in which the $\theta$ parameter is considered as a dynamical field, the axion, which runs to the value zero no matter its initial value. The effective lagrangian describing the axion $a$ and its interaction with the gluons is
$$
L_a=L_{QCD}+L_k(a)+(\theta+\frac{a}{f_a})\frac{\alpha_s}{8\pi}G^a_{\mu\nu}\widetilde{G}^{a\mu\nu},
$$
with $f_a$ the axion constant and $L_{k}(a)$ its kinetic term. 
\begin{figure}[!htb]
\begin{center}
\begin{tabular}{cccccccccccccccc}    

\begin{fmffile}{onegine10} 	
  \fmfframe(6,42)(6,42){ 	
   \begin{fmfgraph*}(220,124) 
    \fmfleft{i}	
    \fmfright{o1,o2}    
    \fmflabel{$a$}{i}
    \fmflabel{$q_1$}{o1}
    \fmflabel{$q_2$}{o2}
    \fmf{scalar}{i,iv3}
   \fmf{fermion}{o1,ov1}
   \fmf{fermion}{o2,ov2}
    \fmf{fermion,label=$\psi$}{iv1,iv2,iv3,iv1}
    \fmf{fermion}{ov1,ov2}
    \fmf{gluon,label=$G$}{iv1,ov1}
    \fmf{gluon,label=$G$}{iv2,ov2}
  	\fmffixed{(0,.7h)}{iv1,iv2} \fmffixed{(0,.7h)}{ov1,ov2}
     \end{fmfgraph*}
  }
\end{fmffile}
\end{tabular}
\caption{\textsf{The diagram giving an effective coupling $a G\widetilde{G}$.}}\label{fey121}
\end{center}
\end{figure}

This corresponds to an effective $\overline{\theta}$ term with
$$
\overline{\theta}=\theta+\frac{a}{f_a}.
$$
By shifting the field $a\to a-f_a\theta$ the $\theta$ parameter can be discarded. Thus the theory will be CP invariant if there is something that
forces the axion to take the value $a=0$. This is precisely what happens, since the axion is under the influence of an effective quantum potential $V(a)$ due to the effect of the quarks and gluons inside the Feynmann path integral whose minimum is $a=0$. By definition this potential is given by
$$
\exp(-\int d^4x V(a))=\int [DA_\mu]\prod_{i}[Dq_i][D\overline{q}_i]\exp\bigg[-\int d^4x\bigg( L_{QCD}+\frac{\alpha_s}{8\pi f_a}aG^a_{\mu\nu}\widetilde{G}^{a\mu\nu}\bigg)\bigg],
$$
and its explicit expression has been presented for instance in \cite{axion8}
\be\lb{poto}
V(a)\sim f^2_{\pi}m^2_{\pi}\bigg[1-\cos(\frac{a}{f_a})\bigg].
\ee
The last formula implies that the minima is at $a=0$ and this solves the CP problem. Additionally it implies that the axion possess a mass of the order
\be\lb{rev}
m_a\sim \frac{f_{\pi}m_{\pi}}{f_{a}}.
\ee
Here $f_{\pi}$ and $m_{\pi}$ are the pion coupling constant and its mass respectively. 

The results given above take into account the color interaction. But this should be supplemented with the CP violating terms of the weak interaction, and this results in a very tiny but non zero value of $\theta$.

There are several axion scenarios discussed in the literature \cite{axion11}-\cite{axion9}. In some of them
the axion does not interact directly with the ordinary quarks, and the coupling $a G \widetilde{G}$ is interpreted of an effective interaction. For instance the diagram of the figure 1, for which the triangle is composed by a new heavy quark, gives rise of an effective interaction of the form $f^{-1}_{a}a G \widetilde{G}$ in which the axion parameter $f_a$ is a function of the mass of this new quark.

Let us consider, following \cite{kimfis} and \cite{axion9}, the addition of the following lagrangian
\be\lb{ax}
L_{add}=i\overline{\psi}\widehat{D}\psi-(\delta\varphi \overline{\psi}_R \psi_L+\delta^\ast\varphi^\ast \overline{\psi}_L \psi_R)+(\partial_{\mu}\varphi^\ast)(\partial_{\mu}\varphi)
+m^2\varphi^\ast\varphi-\lambda(\varphi^\ast\varphi)^2,
\ee
to the ordinary QCD. The parameters $\lambda$, $m$ and $\delta$ are constants to be determined and the term $i\overline{\psi}\widehat{D}\psi$ includes the kinetic energy of the new quark $\psi$ and its coupling with the gluons. Since the new scalar field acquires a non zero expectation value $\mid<\varphi>\mid=\varphi_0=m/\sqrt{2\lambda}$, it acquires a mass $m/\sqrt{2\lambda}$. In addition there is a massless pseudoscalar $a$, defined by 
\be\lb{aha}
\varphi=(\varphi_0+\rho) \exp(\frac{ia}{\varphi_0\sqrt{2}}).
\ee
The field $\rho$ describe the radial excitations and $a$ the angular ones. This pseudoscalar is identified with the axion and it is the Goldstone boson associated to the breaking of the $U(1)$ symmetry
\be\lb{mic}
\psi\to e^{i\gamma_5 \alpha}\psi,\qquad \varphi\to e^{-2i\alpha}\varphi,
\ee
of the lagrangian (\ref{ax}). This field does not interact directly with the light quarks and with the gluons, but acquires an effective interaction with the last due to the diagram of the figure 1. The resulting interaction is
$$
\frac{\alpha_s^2}{8\pi\sqrt{2}\varphi_0}a G^a_{\mu\nu}\widetilde{G}^{a\mu\nu}.
$$
and from here it follows that $f_a=\sqrt{2}\varphi_0$. Since the mass of the heavy quark is proportional to $f_a$, the heavier the quark is, the lighter the axion will be.

Whether or not the axion mechanism solves the CP problem in QCD is an open question, but this scenario will be helpful for constructing our model.

\section{Proposed scenario}

The model we will propose is a toy one, with the particularity that it gives a dynamical interpretation of the density of energy of the present universe, and simultaneously contains long lived super-heavy particles, which may be interpreted as super-heavy components of the dark matter. Our model contains a hidden sector with a super-heavy Higgs like boson $\widetilde{H}$ and an axion pseudoscalar $\widetilde{a}$, the last acquires a small mass due to a mixing with the ordinary pion. As we will see, the boson $\widetilde{H}$ is long lived and may be present at our era, while the axion $\widetilde{a}$ will induce an effective energy density of the order of the measured in the present universe. Our approach is a quintessence like model, in which it is assumed that the energy density is zero except for the contribution of the hidden axion $\widetilde{a}$. 

\subsection{The hidden axion and the vacuum energy density of the universe} 

The following discussion will be extremely important for constructing our model. As is well known in classical cosmology, the Hubble parameter is related to the critical density $\rho_c$ and the Newton constant $G_N$ by the Friedmann equation
\be\lb{hub}
H^2=\frac{8\pi}{3}G_N \rho_c.
\ee
This is a classical equation. Besides, the actual Hubble constant satisfy the following approximate numerical relation with the pion mass
\be\lb{hubact}
H_0\simeq G_N m_{\pi}^3.
\ee
This is written in natural units, and the right hand should be multiplied by a factor $c^2/\hbar^2$ for other unit systems. The relation (\ref{hubact}) may be interpreted as a mere coincidence. We will not adopt this vision, instead we will interpret it as a signal of new quantum physics. The left hand side of (\ref{hubact}) is the result of experimental cosmological observations, but the right hand involves two of the fundamentals scales of physics, related to the strong and the gravitational interactions. We will adopt a dynamical interpretation of this intriguing relation by modelling the vacuum energy with a extremely light particle $\widetilde{a}$ with mass $m_{\widetilde{a}}\simeq H_0$. The equation (\ref{hubact}) shows that
\be\lb{hub3}
m_{\widetilde{a}}\sim \frac{m_{\pi}^3}{M_{Pl}^2}\sim 3.10^{-32}eV.
\ee
The relation (\ref{hub3}) relates the mass of the pseudoscalar $\widetilde{a}$ with the pion mass $m_{\pi}$, and this resembles formula (\ref{rev}) which appears in several axion scenarios. This suggests that $\widetilde{a}$ itself may be interpreted as an axion pseudoscalar related to a symmetry breaking mechanism in a hidden sector. This idea is implemented for instance in \cite{frieman} and our goal is to construct an scenario which realize this idea in explicit form. Inspired by \cite{frieman}, we consider a sector which an interaction with ordinary matter weak enough in order to do not be detected by the current accelerator technology. We may refer to this as a hidden sector. Let us assume that this sector contains two fermion particles with spin $1/2$, which we name as hidden neutrino $\widetilde{\nu}$ and hidden electron $\widetilde{e}$. We assume that the right component of the hidden neutrino is coupled to an scalar field $\Phi$, in analogy fashion as in (\ref{ax}). The coupling we choose is given by a Majorana term of the form
\be\lb{fu}
L=\Phi \psi_{\widetilde{\nu}_R}^T C \psi_{\widetilde{\nu}_R},
\ee
$C$ being a representation of the charge conjugation matrix. Additionally, we assume that this sector is invariant under a $SU(2)_L$ gauge symmetry, which relates the left components $\widetilde{\nu}_L$ and $\widetilde{e}_L$ of the hidden fermions. The scalar field $\Phi$ does not couple to these components, therefore it does not have $SU(2)_L$ quantum numbers. The presence of the gauge symmetry implies that there appear three vector bosons, which we will denote as $\widetilde{W}_i$ with $i=1,2,3$. We assume that the interaction between the ordinary sector and the hidden one is mediated by these bosons. Therefore the standard model particles are assumed to have charges corresponding to this hidden gauge symmetry. The lagrangian constructed in these terms resembles the one of section 2, except that the gauge group is not the same and instead of adding a new fermion, we have divided our multiplet into left and right components. In any case, the lagrangian for this hidden sector is invariant under a $U(1)$ chiral symmetry and the associated current $\widetilde{J}_5$ has an anomaly of the form (\ref{anomu}) but with the coupling constant $g_{\widetilde{W}}$ playing the role of $g_{c}$. The hidden axion $\widetilde{a}$ is identified through the equality
$$
\Phi=(f_{\widetilde{a}}+\widetilde{\rho})e^{i\frac{\widetilde{a}}{f_{\widetilde{a}}}},
$$
which is the analogous of the relation (\ref{aha}) for our scenario. By arguments completely analogous to the ones in section 2, it follows that this pseudoscalar is under the influence of an effective potential
$$
V(\widetilde{a})\sim M^4\bigg[1-\cos\bigg(\frac{\widetilde{a}}{f_{\widetilde{a}}}\bigg)\bigg].
$$
analogous to (\ref{poto}). The axion mass is given by
\be\lb{isi}
m_{\widetilde{a}}\sim\frac{M^4}{f^2_{\widetilde{a}}}.
\ee
Our proposal is that the interaction between the hidden and ordinary sector is very weak, even of the gravitational order \cite{Estrada}\footnote{In fact the idea of an interaction of gravitational order between the hidden and ordinary sector has been developed in \cite{gravhidd} in the context of D-branes.}.  We chose $f_a$ to be of the order of the Planck mass 
\be\lb{acc}f_{\widetilde{a}}\sim 10^{19}GeV.
\ee By taking into account the value $m_{\widetilde{a}}\sim 10^{-32}$eV found above it follows from (\ref{isi}) that the mass scale in the potential is $M\sim (10^{-2}- 10^{-3})$eV. This small value implies that the potential is nearly flat. The formula (\ref{hub3}) can be rewritten in terms of $f_{\widetilde{a}}$ as
\be\lb{cer}
m_{\widetilde{a}}\sim \frac{m_{\pi}}{M_{Pl}}\frac{f_{\pi}}{f_{\widetilde{a}}}m_{\pi}.
\ee
The last relation resembles(\ref{rev}) but corrected by a factor $m_{\pi}/M_{Pl}$.
We interpret this correction as coming from switching the strong interaction, for which the pion is the characteristic scale, to the hidden one, whose characteristic scale is the Planck one. Being this axion so light,
the mass of right component of the hidden neutrino $\widetilde{\nu}_R$ happens to be very large. In fact it is given by $m_{\widetilde{\nu}_R}\simeq f_{\widetilde{a}}$, the numerical result is
\be\lb{injv}
m_{\widetilde{\nu}_R}\sim 10^{19}GeV,
\ee
and it is of the order of the Planck mass.

\subsection{A Higgs mechanism in the hidden sector}

We have constructed a toy model in which the density of energy of the universe is interpreted in terms of a symmetry breaking mechanism in a hidden sector. Our next task is to cast long lived super-heavy particles in this sector, which may act as a super-heavy component of dark matter at the present times.

We assume that the masses of the hidden particles are given by a Higgs mechanism, and we are going to fix the scales of the model in such a way that the hidden Higgs  $\widetilde{H}$ is super massive and long lived. But before to do that, some subtle details should be discussed first. The simplest decay channel for $\widetilde{H}$ would be the the diagram of the figure 2, in which the decaying products may be the hidden electron $\widetilde{e}$ or the hidden neutrino $\widetilde{\nu}$. The resulting amplitude is calculated in all the standard textbooks of QFT and it follows that the interaction must be weaker than gravity to obtain the desired mean life. In order to avoid to go beyond the Planck scale, let us consider instead that $\widetilde{H}$ is lighter than the other particles of the model, and does not have a direct coupling with the standard model particles, but acquires effective interaction with them one due to a diagram of the type of the figure 3. This is inspired in the diagram of figure 1 with the scalar $\widetilde{H}$ replacing the axion $a$, and it is assumed that the bosons mediating the interaction between the hidden and ordinary particles the hidden bosons $\widetilde{W}_i$ discussed in the previous subsection.

\begin{figure}[!htb]
\begin{center}
\begin{tabular}{cccccccccccccccc}    

\begin{fmffile}{onegine0} 	
  \fmfframe(6,42)(6,42){ 	
   \begin{fmfgraph*}(220,124) 
    \fmfleft{i}	
    \fmfright{o1,o2}    
    \fmflabel{$X$}{i}
    \fmflabel{$\nu_1$}{o1}
    \fmflabel{$\nu_2$}{o2}
    \fmf{scalar}{i,iv1}
   \fmf{fermion}{iv1,o1}
   \fmf{fermion}{iv1,o2}
     \end{fmfgraph*}}
\end{fmffile}
\end{tabular}
\caption{\textsf{A diagram giving a short-lived scalar particle $X$ which decay into two fermions $\nu_i$.}}\label{fey211}
\end{center}
\end{figure}

Let us focus our attention in the diagram of the figure 3. The triangle is composed by a fermion which should be heavier than $\widetilde{H}$, otherwise the main decay channel for $\widetilde{H}$ will be diagram 2, which is what we would like to avoid. Thus the hidden Higgs $\widetilde{H}$ should be lighter than the fermions of the model $\widetilde{\nu}$ or $\widetilde{e}$, for this reason these fermions are also super-heavy. By denoting the minimum of the Higgs doublet $\widetilde{H}$ as
$$
\mid<\widetilde{H}>\mid=\bigg(\begin{array}{c}
                   0 \\
                   v
                 \end{array}\bigg),
$$
the standard Higgs mechanism implies that the vector bosons and the hidden Higgs acquire a mass $2m_{\widetilde{W}}=g v$ and \be\lb{musa}m_{\widetilde{H}}=2\lambda^{\frac{1}{2}}v, \ee
respectively. We should remark that $v$ and the coupling constant $\lambda$ appearing in the potential $V(\widetilde{H})$ for the hidden Higgs are free parameters. This means that, at this stage, one can give any mass to these particles. Now, if we require the diagram of the figure 3 to be the main decay channel, the vertex of the figure 2 is needed.  This vertex may be induced by the following Dirac coupling between the hidden Higgs doublet $\widetilde{H}$ with the hidden electron and the hidden neutrino 
$$
f_{\widetilde{e}} \overline{l}_L \widetilde{H} \psi_{\widetilde{e}_R}=f_{\widetilde{e}} \overline{\psi}_{\widetilde{e}_L} \frac{v}{\sqrt{2}} \psi_{\widetilde{e}_R}+f_{\widetilde{e}} \overline{\psi}_{\widetilde{e}_L} \widetilde{H} \psi_{\widetilde{e}_R},
$$
\be\lb{dira}
f_{\widetilde{\nu}} \overline{l}_L \widetilde{H}^c \psi_{\widetilde{\nu}_R}=f_{\widetilde{\nu}}  \overline{\psi}_{\widetilde{\nu}_L} \frac{v}{\sqrt{2}} \psi_{\widetilde{\nu}_R}+f_{\widetilde{\nu}}  \overline{\psi}_{\widetilde{\nu}_L} \widetilde{H} \psi_{\widetilde{\nu}_R}.
\ee
Here $l_L$ represents the doublet composed by the left components of the hidden electron and hidden neutrino and $f_{e,\nu}$ are new free parameters.
The couplings (\ref{dira}) induce mass terms for the hidden fermions $\widetilde{\nu}$ and $\widetilde{e}$, which generic masses \be\lb{mt}m_{\widetilde{\nu}, \widetilde{e}}=v f_{\widetilde{\nu}, \widetilde{e}}.\ee This represents the mass of the hidden electron, but the mass of the hidden neutrino is more tricky since there are two couplings giving mass terms for this particle namely, (\ref{dira}) and (\ref{fu}). Therefore the hidden neutrino acquires a non trivial mass matrix given by
\be\lb{mus}
M_{\widetilde{\nu}}=\bigg(\begin{array}{cc}
                      0 & m_{\widetilde{\nu}} \\
                      m_{\widetilde{\nu}}  & m_{\widetilde{\nu}_R}
                    \end{array}\bigg),\qquad m_{\widetilde{\nu}_R}\simeq f_{\widetilde{a}},
\ee
We should require that none of the eigenvalues of this mass matrix is smaller than the Higgs mass, otherwise the main decay will be the diagram of the figure 2, which is what is to be avoided. In addition, given that (\ref{injv}) implies that $m_{\widetilde{\nu}_R}$  is of the order of the Planck mass, we may assume that $m_\nu<< m_{\widetilde{\nu}_R}$. Under this conditions the approximate eigenvalues of the mass matrix are
\be\lb{ein}
M_1\simeq m_{\widetilde{\nu}_R},\qquad M_2\simeq\frac{m^2_{\widetilde{\nu}}}{ m_{\widetilde{\nu}_R} }
\ee
with the eigenvectors
$$
\psi_{\widetilde{\nu}}^1=\frac{1}{\sqrt{1+\bigg(\frac{m_{\widetilde{\nu}_R}}{ m_{\widetilde{\nu}}}\bigg)^2}}\psi_{\widetilde{\nu}_L}+\frac{\bigg(\frac{m_{\widetilde{\nu}_R}}{ m_{\widetilde{\nu}}}\bigg)}{\sqrt{1+\bigg(\frac{m_{\widetilde{\nu}_R}}{ m_{\widetilde{\nu}}}\bigg)^2}}\psi_{\widetilde{\nu}_R},
$$
$$
\psi_{\widetilde{\nu}}^2=\frac{1}{\sqrt{1+\bigg(\frac{ m_{\widetilde{\nu}}}{m_{\widetilde{\nu}_R}}\bigg)^2}}\psi_{\widetilde{\nu}_L}-\frac{\bigg(\frac{ m_{\widetilde{\nu}}}{m_{\widetilde{\nu}_R}}\bigg)}{\sqrt{1+\bigg(\frac{ m_{\widetilde{\nu}}}{m_{\widetilde{\nu}_R}}\bigg)^2}}\psi_{\widetilde{\nu}_R}.
$$
It follows from the last expressions that if $m_\nu<< m_{\widetilde{\nu}_R}$, as we are assuming, then $\psi_{\widetilde{\nu}}^1\simeq\psi_{\widetilde{\nu}_R}$
and $\psi_{\widetilde{\nu}}^2\simeq\psi_{\widetilde{\nu}_L}$. This is an example of a see-saw mechanism.

\begin{figure}[!htb]
\begin{center}
\begin{tabular}{cccccccccccccccc}    

\begin{fmffile}{onegine9} 	
  \fmfframe(6,42)(6,42){ 	
   \begin{fmfgraph*}(220,124) 
    \fmfleft{i}	
    \fmfright{o1,o2}    
    \fmflabel{$X$}{i}
    \fmflabel{$q_1$}{o1}
    \fmflabel{$q_2$}{o2}
    \fmf{scalar}{i,iv3}
   \fmf{fermion}{o1,ov1}
   \fmf{fermion}{o2,ov2}
    \fmf{fermion,label=$\nu$}{iv1,iv2,iv3,iv1}
    \fmf{fermion}{ov1,ov2}
    \fmf{gluon,label=$\widetilde{W}$}{iv1,ov1}
    \fmf{gluon,label=$\widetilde{W}$}{iv2,ov2}
  \fmffixed{(0,.7h)}{iv1,iv2} \fmffixed{(0,.7h)}{ov1,ov2}
     \end{fmfgraph*}
  }
\end{fmffile}
\end{tabular}
\caption{\textsf{The main decay channel of the hidden Higgs. It resembles figure 1 with the Higgs replacing the axion and the quark $\psi$ replaced by ultra-heavy fermions in the triangle. The gauge fields replacing the gluons correspond to a very weak interaction with the ordinary matter, which is represented by the output fermion lines of this diagram.}}\label{fey432}
\end{center}
\end{figure}

The next task is give values to the four free parameters on the model namely $v$, $\lambda$ and $f_{e,\nu}$.  We took $f_{\widetilde{e},\widetilde{\nu}}\simeq g\simeq 10^{-1}$ and the mass of the vector bosons of the order of the GUT scale, which gives $v\simeq 10^{17}$GeV. This expectation value gives \be\lb{meh}m_{\widetilde{e},\widetilde{\nu}}\sim10^{16}GeV,\ee also of the order of the GUT scale. Since the left and right components of the electron have the same mass, it is safe to say that the electron has a mass of the order of the GUT scale. But the neutrino has two eigenvalues (\ref{ein}), the first $M_1$ is by (\ref{injv}) of the order of the Planck mass. The other eigenvalue may be calculated by taking into account that $m_{\widetilde{\nu}}\simeq 10^{16}$GeV, the result is \be\lb{ali}M_2\sim 10^{13}GeV.\ee Still the parameter $\lambda$ is not defined, which means that the mass of the hidden Higgs is arbitrary. If the diagram 3 is the main decay channel, this mass should be smaller that the minimal mass of the fermions. This fix $m_{\widetilde{H}}<M_2\sim 10^{13}$GeV and, by taking into account (\ref{musa}), it follows that $\lambda<10^{-3}$. But although $m_{\widetilde{H}}$ is smaller than $M_2$, we are going to require both to be of the same order of magnitude, i.e, $\lambda\simeq 10^{-3}$. This is in order to keep the hidden Higgs as massive as possible.

The previous discussion fix all the parameters of the model except $g_{\widetilde{W}}$. This coupling constant
will be estimated in the next section by requiring that the mean time life of the Higgs to be of the order or even larger than the age of the universe.

\section{Hidden Higgs lifetime}

The leading decay mechanism for the hidden Higgs is the diagram of the figure 3, which corresponds to the following amplitude
$$
{\cal{M}}=\frac{g_1(g_{\widetilde{W}})^2}{2\pi^8}\sum_{ijk}\int\int dk_1^4 dk_2^4 \frac{tr\{[\displaystyle{\not}p_1-\displaystyle{\not}k_1-\displaystyle{\not}k_2+m_{F_i}I]\gamma^\mu [ \displaystyle{\not}p_2-\displaystyle{\not}k_1-\displaystyle{\not}k_2+m_{F_j}I](\displaystyle{\not}k_2+m_{F_k}I)\gamma^\nu\}}{[(p_1-k_1-k_2)^2-m_{F_i}^2][(p_2-k_1-k_2)^2-m_{F_j}^2]}
$$
\be\lb{amp}
\frac{\eta_{\mu d}\eta_{\nu\xi}(\gamma^d)^{mj}(\gamma^{\xi})^{ek}\overline{\psi}_e \psi_m (\displaystyle{\not}k_1+mI)_{jk}}{[k_2^2-m_{F_k}^2][k_1^2-m^2](p_2-k_1)^2(p_1-k_1)^2}.
\ee
Here $m$ can be is a typical mass of a light fermion in the usual standard model and $m_{f_i}$ is the mass of the fermions on the triangle which, as we discussed, can take the values $m_e$, $M_1$ or $M_2$. Also $g_1$ can be $f_{e}$ or $f_{\nu}$ and $g_{\widetilde{W}}$ is the coupling constant of the interaction between the hidden and the ordinary sector. We have discussed in the previous section that this interaction may be of gravitational order, but we are going to make an explicit estimation of its value.

We have neglected the mass of the $\widetilde{W}_i$ bosons in the propagators intentionally and this requires a justification. We are looking for a lower bound for the mean life time or, what is the same, an upper bound for ${\cal{M}}$. The vector bosons appear as internal lines and we expect the probability of the decay to go to zero as their mass goes to infinite, i.e, ${\cal{M}}\to 0$ when $m_{\widetilde{W}}\to \infty$. Reversing the argument we expect that the amplitude ${\cal{M}}$ for $m_{\widetilde{W}}\to 0$ gives a life time which is smaller than the real one, and this is the bound we are looking for. Besides, it is more easy to estimate  ${\cal{M}}$ by a dimensional analysis argument in a situation which has less parameters.

Our first step is to calculate the graph as if the internal lines in the triangle were composed by one fermion, and the mixing effect due to the mass matrix (\ref{mus}) will be consider latter on. The minimal mass for the hidden fermions follows from (\ref{ali}) and is $m_f\simeq 10^{13}$GeV. We will choose this mass in the following, since it is the value which gives the largest decay probability. 

At first sight, the integral seems to have a logarithmic divergence. The numerator is an expression of degree four, the denominator is of degree twelve and the integration is in eight variables. But when the trace is expanded it is found that
$$
tr\{[\displaystyle{\not}p_1-\displaystyle{\not}k_1-\displaystyle{\not}k_2+m_{F_i}I]\gamma^\mu [\displaystyle{\not}p_2-\displaystyle{\not}k_1-\displaystyle{\not}k_2+m_{F_j}I](\displaystyle{\not}k_2+m_{F_k}I)\gamma^\nu\}=(\displaystyle{\not}p_1
-\displaystyle{\not}k_1-\displaystyle{\not}k_2)_{\xi}(\displaystyle{\not}p_2-\displaystyle{\not}k_1-\displaystyle{\not}k_2)_{\eta}
$$
$$
\times tr(\gamma^\xi\gamma^\mu\gamma^\eta\gamma^\delta \gamma^\nu)+m_F (\displaystyle{\not}p_1-\displaystyle{\not}k_1-\displaystyle{\not}k_2)_\xi(\displaystyle{\not}p_1-\displaystyle{\not}k_1-\displaystyle{\not}k_2)_\eta  tr(\gamma^\xi\gamma^\mu\gamma^\eta \gamma^\nu)+...
$$
where the dots denote terms with lower powers in the momentum. The first term of this expansion is zero, since it is multiplied by the trace of a product of an odd number of gamma matrices. This means that the numerator is cubic and not quartic in the external momenta, and the integral is convergent instead of logarithmic divergent. But before to make the estimation of this integral, it will be convenient to compare our situation with some result of the literature.
\\

\textit{Comparison with an axion mechanism due to Kim:} The estimation of the amplitude ${\cal{M}}$ we are going to do require to understand properly the qualitative behaviour of (\ref{amp}) with respect to the mass parameters of the model. Consider first the situation in which the masses $m_f$ corresponding to the fermion triangle take very large values $m_{f}\to \infty$. It is reasonable to expect that the decay probability decrease to zero and therefore the mean life time $\tau$ becomes infinite in this limit. In other words, it is expected that 
\be\lb{con1}
\lim_{m_{f}\to \infty}\tau\to \infty.
\ee
It is also reasonable to expect that when $m_{\widetilde{H}}$ increase, the decay is more probable and the mean life time decreases, that is 
\be\lb{con2}
m_{\widetilde{H}}<m^{'}_{\widetilde{H}},\qquad \Rightarrow,\qquad \tau(m_{\widetilde{H}})>\tau(m^{'}_{\widetilde{H}}).
\ee
Instead, the behaviour of the amplitude with respect to the mass $m$ of the products of the decay is more involved. A decrease of this mass is equivalent to an increase the masses $m_f$ and $m_{\widetilde{H}}$ with their ratio $R=m_{\widetilde{W}}/m_{f}<1$ fixed. The increase of the mass of the Higgs make the life time shorter by (\ref{con2}), while the increase of the triangle mass makes it larger by (\ref{con1}). It is not clear with this argument which of the two effects prevails, if any. Fortunately, we can get a clue by consulting some literature about axions. For instance, the work \cite{kimfis} of Kim considers essentially the same diagram than the one in the figure 3, but with an axion replacing the hidden Higgs, a heavy quark $Q$ replacing our hidden neutrino or hidden electron, and gluons replacing our $\widetilde{W}_i$ bosons. The Kim diagram induce an effective coupling between the decaying axion and the ordinary quarks. The effective coupling the author calculates is schematically
\be\lb{kimon}
g_{eff}\simeq\frac{g_{c}^4}{v} m_q \ln(\frac{m_Q}{m_q}).
\ee
Here $v$ is an expectation value defined in that paper, and $m_q$ is a mass of an ordinary quark, which is the product of the decay. By using the elementary L'Hopital rule it is found that $g_{eff}\to 0$ when the masses $m_q$ goes to zero. Thus the mean life time of the axion becomes infinite in this limit, since there is no coupling and thus no decay. This means that the lighter become the ordinary quarks are, the larger the mean life time of the axion becomes. Note that this is not what a first intuition would suggest, but is the result of the competition between the two effects previously mentioned.

The main difference between the Kim decay described above and the one we are presenting is that the Kim axion is a pseudoscalar while our hidden Higgs is an scalar. We will assume that this difference does not alter the behaviour of the amplitude on the mass $m$ of the products of the decay. Our assumption will be that
\be\lb{con3}
\lim_{m\to 0} \tau\to\infty,
\ee
as in the Kim model.
 
We should stress that we can not borrow the result of the Kim calculation directly, the reason is that the mass of the heavy quarks and the axion mass are not independent in the Kim scenario, in fact both are given in terms of the expectation value $v$. Thus, the Kim calculation essentially involve two parameters, while in our case there are three mass parameters. Despite this technical problem, we will make an independent estimation of the hidden Higgs life time by taking into account the three conditions (\ref{con1}), (\ref{con2}) and (\ref{con3}) and the expression (\ref{kimon}) as a partial guide.
\\

\textit{Estimation of the mean life time of the hidden Higgs:} In order to calculate the decay width $\Gamma$ and the mean lifetime $\tau=\Gamma^{-1}$ of the boson $\widetilde{H}$ we choose a coordinate system in which it is at rest
$$
q=(m_{\widetilde{H}},0,0,0),\qquad p_1=(p_{10},p_{1x},0,0), \qquad p_2=(p_{20},p_{2x},0,0).
$$
The conservation of energy and momenta implies that
$$
p_{1x}=-p_{2x},\qquad p_{1,2 0}^2- p_{1,2 x}^2=m^2,\qquad p_{10}+p_{20}=m_{\widetilde{H}},
$$
and it is not difficult to show that
$$
p_{10}=p_{20}=\frac{m_{\widetilde{H}}}{2}.
$$
The decay width $\Gamma$ is obtained by integrating
$$
d\Gamma=\frac{1}{m_{\widetilde{H}}}\sum_{s_1s_2}|{\cal{M}}|^2\frac{d^3p_1d^3p_2}{(2\pi)^2E_1 E_2}\delta^4(p_1+p_2-q),
$$
with respect to $p_2$ and $p_1$, which gives
$$
\Gamma=\frac{1}{m_{\widetilde{H}}}\int\frac{ d^3p_1\delta(\sqrt{p_1^2+m^2}-\frac{q}{2})}{(\sqrt{p_1^2+m^2})^2}\sum_{s_1s_2}|{\cal{M}}|^2.
$$
By defining the more convenient variable $u=\sqrt{p_1^2+m^2}$, the decay width may be expressed as
$$
\Gamma=\frac{1}{m_{\widetilde{H}}}\int\frac{ du\sqrt{u^2-m^2}}{u}\delta(u-\frac{q}{2})\sum_{s_1s_2}|{\cal{M}}|^2,
$$
and, after integration, it follows that
$$
\Gamma=\frac{\sqrt{m_{\widetilde{H}}^2-m^2}}{m^2_h}\sum_{s_1s_2}|{\cal{M}}|^2\simeq \frac{1}{m_{\widetilde{H}}}\sum_{s_1s_2}|{\cal{M}}|^2.
$$
In the last step we took into account that $m_{\widetilde{H}}>>m$. 

The next task is to calculate explicitly (\ref{amp}), to insert the result into the last
formula after that and to perform the sum over the polarizations of the particles resulting from the decay. This is a complicated
task, and we do not know of the result of the eight dimensional integral explicitly. Therefore we will conform ourselves with an estimation, given as follows. Clearly ${\cal{M}}^2$ is an expression of the form $[\overline{\psi}^{s_1}(p_1)\gamma_{\xi}(k_1+mI)\gamma_{\mu}\psi^{s_2}(p_2)]$. Since this expression is a finally a number and the gamma matrices are hermitian we have that
$$
[\overline{\psi}^{s_1}(p_1)\gamma_{\xi}(\displaystyle{\not}k_1+mI)\gamma_{\mu}\psi^{s_2}(p_2)]^\ast=
[\overline{\psi}^{s_1}(p_1)\gamma_{\xi}(\displaystyle{\not}k_1+mI)\gamma_{\mu}\psi^{s_2}(p_2)]^{\dagger}=
\overline{\psi}^{s_2}(p_2)\gamma_{\mu}(\displaystyle{\not}k_1+mI)\gamma_{\xi}\psi^{s_1}(p_1).
$$
With a help of the last expression and, by taking into account the equality
$$
\sum_{s_1}[\overline{\psi}^{s_1}(p_1)]^A[\psi^{s_1}(p_1)]^B=(\displaystyle{\not}p_1+mI)^{AB},
$$
it may be shown that $\sum_{s_1, s_2}|{\cal{M}}|^2$ is an expression which contains $p_{1,2}^2=m^2$ and $p_1\cdot p_2\sim m_{\widetilde{H}}^2$. We will replace all these expressions by the largest mass scale, namely $m_{\widetilde{H}}^2$, since we are interested to obtain a lower bound of a lifetime or, what is the same, an upper bound
for the decay width $\Gamma$. By replacing all the expressions by the largest mass scale it follows that
$$
|{\cal{M}}|^2\simeq\alpha_1\alpha_{\widetilde{W}}^4 m_{\widetilde{H}}^2 I^2(m,m_f,m_{\widetilde{H}})
$$
with $4\pi\alpha_{1,\widetilde{W}}=g_{1,\widetilde{W}}^2$ and $I(m,m_f,m_{\widetilde{H}})$  a complicated eight dimensional integral with respect to the internal variables $k_1$ and $k_2$. Then
\be\lb{bono}
\Gamma\simeq\alpha_1\alpha_{\widetilde{W}}^4 m_{\widetilde{H}} F(m,m_f,m_{\widetilde{H}}),
\ee
with $F(m,m_f,m_{\widetilde{H}})=I^2(m,m_f,m_{\widetilde{H}})$. Since, in natural units, $\Gamma$ has energy dimensions and $m_{\widetilde{H}}$ is multiplying the whole expression, the function $F(m,m_f,m_{\widetilde{H}})$ should be dimensionless. This means that 
$$
F(m, m_{\widetilde{H}}, m_f)=f(x_1, x_2),\qquad x_1=\frac{m}{m_f},\qquad x_2=\frac{m}{m_{\widetilde{H}}}.
$$
Furthermore, the conditions (\ref{con1}) and (\ref{con3}) imply that
$$
\lim_{x_1\to 0} f(x_1, x_2)\to 0.
$$
Now let us take a look to the Kim formula (\ref{kimon}). The decay rate for the Kim axion is proportional to $g^2_{eff}$, and thus proportional
to $m^2\log^2(x_1)$. We expect this to occur in our case. We postulate that
$$
f(x_1, x_2)=g(\frac{m}{m_{\widetilde{H}}},\frac{m}{m_f})\log^2(\frac{m_f}{m}).
$$
with $g(x_1, x_2)$ an unknown function of the dimensionless variables. Since we know that in the limit $m_f\to 0$ the variable $x_1\to 0$ and the decay rate should go to zero by (\ref{con1}), it follows that $g(x_1, x_2)\to 0$ faster than $\log^{-2}(x_1)$. Let us assume for the moment that this function is analytic with respect to $x_2$, below we will relax this assumption. Since $x_1<<1$ we may perform a Taylor expansion for $g(x_1, x_2)$ in the last formula to obtain 
$$
f(x_1, x_2)\simeq h(\frac{m}{m_{\widetilde{H}}})\frac{m}{m_f}\log^2(\frac{m_f}{m}).
$$
In these terms it follows that (\ref{bono}) is
\be\lb{bono2}
\Gamma\simeq\alpha_1\alpha_{\widetilde{W}}^4 m_{\widetilde{H}}\; h(\frac{m}{m_{\widetilde{H}}})\;\frac{m}{m_f}\log^2(\frac{m_f}{m}).
\ee
Still, we can get more information about the new function $h(x_2)$. As we discussed in (\ref{con1})-(\ref{con3}), the limit taking $m\to0$ is qualitatively equivalent to take the mass $m_{\widetilde{H}}$ and $m_f$ to infinite, but with their ratio fixed $m_f/m_{\widetilde{H}}=R>1$. The condition (\ref{con3}) implies that in this limit the life time goes to infinite and thus (\ref{bono2}) should go to zero. By putting $m_f=R m_{\widetilde{H}}$ in (\ref{bono2}) it follows that the expected limit is
\be\lb{expect}
\lim_{m_{\widetilde{H}}\to\infty} \frac{m}{R}h(\frac{m}{m_{h}}) \log^2(\frac{R m_{\widetilde{H}}}{m})\to 0.
\ee
This means that $h(x) \to 0$ when $x\to 0$ faster than $\log^{-2}(x)$. If we assume this function to be also analytic in the small variable $x_2$ the we 
can approximate (\ref{bono2}) by making a Taylor expansion, whose result is
\be\lb{akw}
\Gamma\simeq c \alpha_1\alpha_{\widetilde{W}}^4 \frac{m^2}{m_f}\log(\frac{m_f}{m}),
\ee
with $c= h^{'}(0)$. Note that we obtained the dependence on $m^2$ we expected. Nevertheless, the statement that $h(x_1, x_2)$ is analytic in $x_2$ made above does not seems satisfactory, since the resulting decay rate (\ref{akw}) obtained under this assumption does not depend on the mass $m_{\widetilde{H}}$ of the decaying particle. If we relax the analyticity condition but still insist that the result should depend on $m^2$, then it may be reasonable to state that \be\lb{rison}
g(x_1, x_2)=c x_1^\alpha x_2^\beta=\frac{c m^2}{m^{\alpha}_{\widetilde{H}} m_f^{\beta}},\ee
with $\alpha+\beta=2$ since $g(x_1, x_2)$ should be dimensionless. Note that $\alpha=\beta=1$ is the case we already considered. For other values, we will have that $\alpha<1$ or $\beta<1$, and $g(x_1,x_2)$ will not be analytic in the corresponding variable $x_1$ or $x_2$. 
With the assumption (\ref{rison}) the decay rate (\ref{bono}) becomes
\be\lb{akw2}
\Gamma\simeq c \alpha_1\alpha_{\widetilde{W}}^4 m_{\widetilde{H}}\frac{m^2}{m^{\alpha}_{\widetilde{H}}m^{\beta}_f}\log^2(\frac{m_f}{m}).
\ee
We expect the decay rate to grow when the mass of the decaying particle $m_{\widetilde{H}}$ increases, and thus we expect that $\alpha<1$. But we do not know the exact value of the powers $\alpha$ and $\beta$. Nevertheless, since we are assuming that $m_{f}$ and $m_{\widetilde{H}}$ are of the same order of magnitude we can put $m_f\simeq m_{\widetilde{H}}$ in (\ref{akw2}) and, by taking into account that $\alpha+\beta=2$, if follows that (\ref{akw2}) is essentially the same (\ref{akw}) in our approximation. We do not have real control on the value of the slope $c$, but we suppose that it do not take values much larger than the unity. 

Having obtained reasonable formulas for the decay rate, the next task is to evaluate it numerically. From the previous section we have that $m_{f}\sim 10^{13}GeV$ and that $f_{\widetilde{e},\widetilde{\nu}}\simeq 10^{-1}$, which implies that $\alpha_1\sim 10^{-2}$. For the mass $m$ we choose a characteristic value for the particles of the standard model namely
$ m\sim 1 MeV $. With these values, and by choosing $\alpha_{\widetilde{W}}\sim 10^{-6}$, it follows from (\ref{akw}) or (\ref{akw2})
\be\lb{tivi}
\Gamma\sim 10^{-34} eV,
\ee
which corresponds to a lifetime \be\lb{lt}\tau\sim 10^{11} yrs.\ee This is already of the order of the estimated age
of the universe.

In the previous calculation we have assumed that only one type of fermion is in the triangle. If we consider the presence of three fermions with different masses, then the dimensional argument given previously does not take place for $F$. But it should be true that
$$
I_{m max}<I<I_{m min},
$$
with $I_{m max}$ and $I_{m min}$ the integrals corresponding to a triangle with the heavier and with the lightest eigenstates of mass. Let us consider for instance the component which is conformed by the left component of the neutrino. Since this component is not an eigenstate of mass, in fact $\nu_L\sim \nu_2-10^{-3}\nu_1$, one has to introduce a modified propagator $\frac{1}{p-m_{\nu_2}}-\frac{10^{-3}}{p-m_{\nu_1}}$ in the calculation of $M$, which gives
$$
I_{\nu_L}=I_{1,1,1}-3.10^{-3}I_{1,1,2}+3.10^{-6}I_{2,2,1}-10^{-9}I_{2,2,2},
$$
where the indices indicate the eigenstates composing the triangle. Since the state with lower mass is $\nu_2$ it follows that
$$
I_{\nu_L}<(1+3.10^{-3}+3.10^{-6}+10^{-9})I_{2,2,2}\simeq I_{2,2,2}.
$$
Therefore
$$
\Gamma_{\nu_L}<\Gamma_{2,2,2}\simeq 10^{-33} eV,
$$
and we conclude that this mixing effect does not spoil our estimation.

\section{Discussion and open perspectives}

In the present work we constructed a hypothetical scenario describing the actual vacuum energy density of the universe in terms of the potential energy of an axion pseudoscalar and, simultaneously, giving rise to an stable super heavy particle, the hidden Higgs, with a mass of the order of $10^{13}$GeV. This unifying property is, in our opinion, the most interesting feature of our scenario. It important to emphasize that our estimation of the decay rate of the hidde Higgs is based on a coupling constant value $\alpha_{\widetilde{W}}\simeq 10^{-6}$  for the interaction between the hidden and the ordinary sector, which is the order of the weak interaction. On the other hand, we have argued that the real interaction must be extremely weak, of the order of the gravitational interaction, if the axion energy density is required to reproduce the one for the actual universe. Therefore the mean life time of our hidden Higgs may several orders of magnitude larger than the age of the universe. In addition we have neglected the masses of the hidden vector bosons in our estimation, which are of the order of the GUT scale. It is likely that the presence of this masses decrease the probability of the decay, and this is other effect which makes the hidden Higgs mean life time even larger.

It should be stressed that there is a line of though in which it is assumed that there exist super heavy components of  the dark matter. The particles composing our hidden sector can be considered as "wimpzillas" in the terminology of \cite{riotto}, that is, weakly interacting matter with masses of the order of the GUT scale. But, different from \cite{riotto2}, we are not considering transplanckian physics. In any case, the analysis made in the present work does not prove that "wimpzillas" exist in nature. Our work is focused in characterizing the mechanisms which, if these hypothetical particles do exist, guarantees their stability.

We would like to remark that there exist some scenarios in the literature giving rise to stable super heavy particles, but these particles are not identified as a hidden Higgs.  Besides, the stability arguments of those references are different than the ones presented here. For instance, the reference \cite{yanagida} present some models with stable super heavy particles, in which the stability is guaranteed by a symmetry protection mechanism involving discrete gauge symmetries, which forbids their decay into the particles of the standard model. This idea was implemented also in \cite{tortoci} to study UHECR anisotropy at low energies. In our scenario instead, the stability is insured by the weakness of the interaction between the hidden and ordinary sector, and by an specific decay mechanism inspired by axion models. In addition there exist scenarios with hidden Higgs such as \cite{wicax}-\cite{Higgs6}, but the masses of the hidden Higgs we are presenting is several orders of magnitude heavier than those ones. In addition there exist scenarios such as \cite{overlap} which contains super heavy dark matter and quintessence field, but in these scenarios the quintessence is originated by decays of super heavy dark matter, and is not interpreted in terms of a symmetry breaking mechanism as we did.
We are not trying to invalidate these scenarios when making this comparison. We are just emphasizing that, to the best of our knowledge, our approach has not been presented in the literature before.

The presented model, as the majority of the quintessence models, assumes that the vacuum density energy is zero except for the contribution of the axion. In this model the vacuum energy is a temporary effect which disappear when the axion reach the minimum of the potential. It may be an interesting task to find scenarios which describes the smallness of the vacuum energy in terms of cancellation mechanisms, and simultaneously giving rise to stable super heavy particles. The main technical problem that we see is that the cancellation mechanisms we know do not involve scalar or pseudoscalar particles, and thus it is not straightforward to reproduce the arguments presented here. In any case, to realize this scenario is an interesting task that deserves to be investigated further.
\\

{\bf Acknowledgements:}  This work arise due to some personal collaboration i had with Luis Masperi. Unfortunately he passed away at middle stages
of the collaboration. His decease saddened all who worked with him. He was an outstanding
personality with a genuine interest and knowledge about science and with a strong compromise in peace matters. Although i do not know if he will agree or not with this version of our initial project, this work is dedicated to him. The author is supported by CONICET (Argentina) and by the ANPCyT grant PICT-2007-00849.
\\


\begin{thebibliography}{99}
\bibitem{Carroll} S. Carroll, W. H. Press and E. Turner, Annu. Rev. Astron. Astrophys. 30 (1992) 499.
\bibitem{Dolgov} A. Dolgov "The Problem of Vacuum Energy and Cosmology" astro-ph/9708045.
\bibitem{Dolgov2} A. Dolgov. Cosmology and New Physics" hep-ph/0606230.
\bibitem{Dolgov3} A.D. Dolgov, M. Kawasaki "Realistic Cosmological Model with Dynamical Cancellation of Vacuum Energy" astro-ph/0307442; "Stability of a cosmological model with dynamical cancellation of vacuum energy" astro-ph/0310822.

\bibitem{Eternity} A. Polyakov Nucl. Phys. B 797 (2008) 199; Nucl. Phys. B 834 (2010) 316.
\bibitem{Eternity2} D. Krotov and A. Polyakov Nucl. Phys. B 849 (2011) 410.
\bibitem{Akhmedov} E. Akhmedov and P. Buidovich Phys. Rev. D 78 (2008) 104005.
\bibitem{Marolf} D. Marolf and I. Morrison "The IR Stability of de Sitter QFT: Physical Initial Conditions" arXiv:1104.4343.
\bibitem{quintesence} S. Carroll Phys. Rev. Lett. 81 (1998) 3067.
\bibitem{Witten} E. Witten "The Cosmological Constant From The Viewpoint Of String Theory" hep-ph/0002297.
\bibitem{axion1} R. Peccei and H. Quinn Phys. Rev. Lett. 38 (1977) 1440; Phys. Rev. D 16 (1977) 1791.
\bibitem{quv} J. Kim and H. P. Nilles Physics Letters B 553 (2003) 1.
\bibitem{Wilcox} M. Hertzberg, M. Tegmark and F. Wilczek Phys. Rev. D 78 (2008) 083507.

\bibitem{Harari}  D. Harari and P. Sikvie Phys. Lett. B 195 (1987) 361; D. Harari and F. Mazzitelli Phys. Lett. B 266 (1991) 269.


\bibitem{Salam} A. Salam Nuovo Cimento 5 (1957) 299.
\bibitem{Lee} T. Lee and C. Yang Phys. Rev. 104 (1956) 254.
\bibitem{mirror} L. Okun Phys. Usp. 50 (2007) 380.
\bibitem{hidden33} J. Feng, H. Tu and Hai-Bo Yu JCAP 0810 (2008) 043.
\bibitem{hidden2} J. Espinosa, T. Konstandin, J.M. No and M. Quiros Phys. Rev. D 78 (2008) 123528.


\bibitem{wicax} B. Patt and F. Wilczek "Higgs-field Portal Into Hidden Sectors", hep-ph/0605188.


\bibitem{Higgs1} O. Bertolami and R. Rosenfeld Int. J. Mod. Phys. A23 (2008) 4817.
\bibitem{Higgs2}  J. March-Russell, S. West, D. Cumberbatch and D. Hooper JHEP 0807 (2008) 058.

\bibitem{Higgs3} C. Englert, T. Plehn D. Zerwas and P. Zerwas Phys. Lett. B 703 (2011) 298.
\bibitem{Higgs5} O. Lebedev and H. Lee Eur.Phys.J. C 71 (2011) 1821.
\bibitem{Higgs6} R. Dick, R. Mann and K. Wunderle Nucl. Phys. B 805 (2008) 207.






\bibitem{review} 	V. Ryabov, V. Tsarev and A. Tskhovrebov Physics Uspekhi 51 (2008) 1091.

\bibitem{hidden3} V. Kuzmin and I. Tkachev JETPLett. 68 (1998) 271;	Phys.Rev. D 59 (1999) 123006;  Phys. Rept. 320 (1999) 199.
\bibitem{massiva} E. Kolb, A. Linde and A. Riotto Phys. Rev. Lett. 77 (1996) 4290.

\bibitem{GKZ} K. Greisen Phys. Rev. Lett 16 (1966) 748; G. Zatsepin and V. Kuzmin Pisma. Zh. Ekzp. Teor. Fiz (1966) 114.
\bibitem{topsec} V.Berezinsky and A.Vilenkin Phys. Rev. D 62 (2000) 083512.


\bibitem{parecido} D. H. Chung, E. Kolb, A. Riotto Phys. Rev. Lett. 81 (1998)
4048.
\bibitem{parecido2} T. Asaka, M. Kawasaki and T. Yanagida Phys. Rev. D 60 (1999) 103518.
\bibitem{Dolgui} A. Dolgov Big Bang and Heavy Particles, hep-ph/0411283v1.
\bibitem{Auger} The Pierre Auger Collaboration Science 9 318 5852 (2007) 938.
\bibitem{suheavy} V. Berezinsky, M. Kachelriess and M. Solberg Phys. Rev. D 78 (2008) 123535.


\bibitem{belavin} A. Belavin, A. Polyakov, A. Schwarz and Y. Typkin Phys. Lett. 59B (1975) 85.
\bibitem{thooft} G. ´t Hooft Phys. Rev. Lett. 37 (1976) 8; Phys. Rev. D 14 (1976) 3432.
\bibitem{axion8} J. Kim Phys. Rept. 150 (1986) 1.
\bibitem{axion11}  F. Wilczek Phys. Rev. Lett. 40 (1978) 279.
\bibitem{axion12} S. Weinberg Phys. Rev. Lett. 40 (1978) 223.

\bibitem{axion3} W. Bardeen and S. Tye Phys. Lett B 74 (1978) 229.
\bibitem{axion4} D. Chang and R. Mohapatra Phys. Rev. D 32 (1984) 293.
\bibitem{axion5} M. Dine, W. Fischler and M. Srednicki Phys. Lett B 104 (1981) 199.
\bibitem{axion6} D. Kaplan Nucl. Phys. B (1985) 260.
\bibitem{axion7} W. Bardeen, R. Peccei and T. Yanagida Nucl. Phys. B (1987) 401.
\bibitem{axion7} M. Srednicki Nucl. Phys B 260 (1985) 689.
\bibitem{kimfis} J. Kim Phys. Rev. Lett. 43 (1979) 103.
\bibitem{axion9} M. Shifman, A. Vainstein and V. Zakharov Nucl. Phys. B 166 (1980) 493.

\bibitem{frieman} J. Frieman, C. Hill and R. Watkins Phys. Rev. D 46 (1992) 1226.
\bibitem{Estrada} J. Estrada Vigil and L. Masperi Mod. Phys. Lett A 13 (1998) 423.
\bibitem{gravhidd} Z. Berezhiani and R. Mohapatra Phys. Rev. D52 (1995) 6607; Z. Berezhiani Phys. Lett. B 147 (1998) 287.
\bibitem{overlap} H. Ziaeepour Phys. Rev. D 69 (2004) 063512.

\bibitem{riotto} E. Kolb, D. Chung and A. Riotto "DARK98" Proceedings of the Second International Conference on Dark Matter in Astro and Particle Physics" Edited by H V Klapdor-Kleingrothaus and L. Baudis.
\bibitem{riotto2} E.  Kolb, A. Starobinsky and I. Tkachev JCAP 0707 (2007) 005.


\bibitem{yanagida} K. Hamaguchi, Y. Nomura and T. Yanagida Phys. Rev. D 58 (1998) 103503.
\bibitem{tortoci}R. Aloisio and F. Tortorici Astropart. Phys. 29 (2008) 307.

\end{thebibliography}
\end{document}